\documentclass[11pt]{article}
\usepackage{amsmath,amssymb,amsfonts}
\usepackage[english]{babel}
\makeatletter
\@addtoreset{equation}{section}
\makeatother

\def\nn{\nonumber} \def\bd{\begin{document}} \def\ed{\end{document}}
\def\ds{\documentstyle}
\let\bm=\bibitem
\newcommand{\be}{\begin{equation}}
\newcommand{\ee}{\end{equation}}
\newcommand{\bea}{\setlength\arraycolsep{2pt} \begin{eqnarray}}
\newcommand{\eea}{\end{eqnarray}}
\newcommand{\hoch}[1]{$\, ^{#1}$}
\def\p{\partial}

\title{\large {\bf Mass and angular momentum of black holes in low-energy heterotic string theory}}
\date{}

\author{Jun-Jin Peng\footnote{pengjjph@163.com}  \\ \\
\small \sl School of Physics and Electronic Science, Guizhou Normal University, \\
\small Guiyang, Guizhou 550001, People's Republic of China \\
\small \sl Institute of Technical Physics, SEEE, Wuhan Textile University,\\
\small Wuhan, Hubei 430073, People's Republic of China
}

\begin{document}

\maketitle
\vspace{20pt}

\begin{center}
\textbf{Abstract}
\end{center}
We investigate conserved charges in the low-energy effective field theory describing
heterotic string theory. Starting with a general Lagrangian that consists of a metric, a
scalar field, a vector gauge field, together with a two-form potential, we derive off-shell Noether
potentials of the Lagrangian and generalize the Abbott-Deser-Tekin (ADT) formalism
to the off-shell level by establishing one-to-one correspondence between the ADT potential
and the off-shell Noether potential. It is proved that the off-shell generalized ADT formalism
is conformally invariant. Then we apply the formulation to compute mass and angular momentum
of the four-dimensional Kerr-Sen black hole and the five-dimensional rotating charged black
string in the string frame without a necessity to transform the metrics into the Einstein frame.

\voffset=-.90pt
\vspace{40pt}

\section{Introduction}

In recent decades, much work has been devoted to seeking exact solutions for the low-energy
limit of heterotic string theory and a lot of analytic solutions with gravitational field,
dilatons, Kalb-Ramond field and gauge fields have been found in the
context of such an effective theory since Gibbons and Maeda first constructed the
four-dimensional static, spherically symmetric black hole solution \cite{GiMasol}, which
was also independently found few years later \cite{GaHoStsol}. We only list partial
solutions here. In the framework of the low-energy effective field theory describing
four-dimensional heterotic string theory, several new static and charged black hole solutions
for the Einstein-Maxwell-Kalb-Ramond-dilaton system
were obtained in \cite{SurDasS}. A rotating charged black hole solution, usually named as
Kerr-Sen black hole \cite{KerrSen}, was found by Sen in the year 1992. Its generalization
with NUT parameters in various dimensions was presented
in \cite{Chow}. In higher dimensions, Mahapatra constructed a five-dimensional rotating
charged black string solution that possesses a gravitational field, a scalar field, an antisymmetric
tensor gauge field and a U(1) field in the low-energy effective field theory \cite{5Dblackst}.
Apart from the black string solution, five-dimensional black ring solutions were also
found \cite{Yazadjiev,Elvang}. These solutions are of great importance to analyze
various non-perturbative aspects of string theory. Further investigation on their
properties, such as thermodynamic properties, entropy bounds and the proof of the full
Penrose inequality, requires to provide systematic methods to identify the conserved charges
in the low-energy effective field theory of heterotic string theory.

Noether theorem has played an important role in defining conserved charges in various
theories of gravity. Up to now, various approaches have been proposed to compute the
conserved charges in the gravity theories through the Noether procedure, such as the
covariant phase space approach \cite{IyerWald,IyerWald2} proposed by Wald and Iyer, the
covariant formalism \cite{BarnichB,Barnich,BarnichC,BCintegC} developed by Barnich,
Brandt and Compere (BBC), the Abbott-Deser-Tekin (ADT) formalism
\cite{AbbottD1,AbbottD2,DeserT1,DeserT2}, and several related methods presented in
\cite{Obukhov,Petrov,ACOlea}. In particular, the ADT formalism, which is defined in
terms of the Noether potential got through the linearized perturbation for the
expression of the equation of motion in a fixed background of spacetime, has made some
progress on computation scheme for conserved charges in various gravity theories.
Since the fixed background metric has to satify the equation of motion
in vacuum, the Noether potential in ADT formalism is on-shell. Recently, in Ref. \cite{KimKY},
Kim, Kulkarni and Yi generalized the on-shell Noether potential in the ADT formalism to
off-shell level on basis of the linearized perturbation of the Noether current and potential
in arbitrary background. Further incorporating a single parameter path in the space of solutions
into the formalism along the line of the BBC method \cite{BarnichB,Barnich,BarnichC,BCintegC},
they proposed a quasi-local formulation of the conserved charges in covariant theories of
pure gravity.

The generalized formalism for the quasi-local conserved charges supplies a more efficient way
to compute the ADT conserved charges and it has been extended to various theories of gravity
coupled with matter fields or not. In \cite{CiteHJPY}, Hyun, Jeong, Park and Yi generalized
the formalism in \cite{KimKY} by including the effect from matter fields and they further showed
that conserved charges via the modified ADT formalism are consistent with those by both the
covariant phase space approach \cite{IyerWald,IyerWald2} and the boundary stress tensor method
\cite{Papadim,Holland,BKraus}. In \cite{JJPeng}, we presented an off-shell Noether current
that is different from the one in \cite{KimKY} by the variation of the Bianchi identity for
the expression of the equation of motion. Then we employed the generalized formulation to calculate
the quasi-local conserved charges of black holes in four-dimensional conformal Weyl gravity
and in arbitrary dimensional Einstein-Gauss-Bonnet gravity coupled to Maxwell or nonlinear
electrodynamics in AdS spacetime. Other applications and generalizations of the modified ADT
formalism can be found in \cite{CiteChernS,MBGhass,ABBGCHJ,Wuli,Setare,HyunJP,CiteLifBH}.

In this paper we shall extend the off-shell generalized ADT method developed in
\cite{KimKY,CiteHJPY} to investigate the conserved charge in the low-energy effective field theory
of heterotic string theory by taking into account of the contributions from the gravitational field,
scalar field, U(1) gauge field, together with the two-form field. We start with a general Lagrangian
including these fields and aim to provide another systematic
approach to compute the mass and angular momentum in the theory. The generality of the
Lagrangian we consider ensures that we can directly perform calculations of the conserved
charges in the string frame without a necessity to transform the metrics into the
Einstein frame. All the results coincide with the ones usually achieved in the Einstein frame
since it is proved that the off-shell generalized ADT formalism is conformally invariant.

Our paper is organized as follows. In Sect. \ref{secone}, we derive the off-shell Noether
currents and potentials to present a covariant formalism of the conserved charges for
the low-energy effective field theory of heterotic string theory. We then apply the
formalism of conserved charges to compute mass and angular momentum of the four-dimensional
Kerr-Sen black hole and the five-dimensional rotating charged black string in Sect. \ref{sectwo}
and Sect. \ref{secthree}, respectively. Sect. \ref{secfour} is our conclusions.
\ref{appA} is devoted to calculating the energy of general static charged black holes
in Einstein-Maxwell-Dilaton gravity. In the \ref{appB}, we investigate the conformal properties
of the conserved charges.

\section{General formalism}\label{secone}

In this section, we shall derive the off-shell Noether and ADT potentials to present a
formulation of the conserved charges for the low-energy effective field theory of heterotic
string theory. The massless bosonic part of the effective action for the theory
describes Einstein gravity coupled to certain matter fields. It consists of a metric $g_{\mu \nu}$,
a scalar field $\varphi$, the two-form Kalb-Ramond antisymmetric tensor field $B_{\mu \nu}$,
together with other gauge fields. In the present work, we take into account of only a single
U(1) component $A_\mu$ of the full set of gauge fields and start with a more general Lagrangian
\bea
\mathcal{L}&=&\sqrt{-g}L= \sqrt{-g}(L_{(R)}+L_{(\varphi)}+L_{(A)}+L_{(B)}) \nn \, , \\
L_{(R)} &=& \lambda(\varphi) R \, ,
\quad L_{(\varphi)} = \chi(\varphi)\nabla^\mu \varphi\nabla_\mu \varphi
+V(\varphi)  \, , \quad
L_{(A)} = N(\varphi)F^{\mu\nu}F_{\mu\nu} \nn \, , \\
L_{(B)} &=& P(\varphi)H^{\rho\mu\nu}H_{\rho\mu\nu}\, , \label{GenLagran}
\eea
in $D-$dimensional spacetimes, where the two-form field strength
$F_{\mu\nu}=2\partial_{[\mu}A_{\nu]}$ and the three-form field strength $H_{\rho\mu\nu}$ is
defined through $H_{\rho\mu\nu}=3\partial_{[\rho}B_{\mu\nu]}-3\alpha A_{[\rho}F_{\mu\nu]}$ in
terms of an arbitrary constant $\alpha$. The functions $\lambda$, $\chi$, $V$, $N$ and $P$
only depend on the scalar field $\varphi$ but not its derivatives. The generality of the five
functions endows the Lagrangian (\ref{GenLagran}) with capability to describe the scalar-tensor
theories of gravity, Einstein gravity coupled to a scalar field non-minimally and
Einstein-Maxwell-Dilaton gravity.
In what follows, all the fields of the theory are collectively denoted
by $\varphi^{(i)}=(g_{\mu \nu},\varphi,A_\mu,B_{\mu \nu})$. With help of the expressions for
the equations of motion, $\mathcal{E}_{(i)}=\big(\mathcal{E}_{(R)}^{\mu\nu}, \mathcal{E}_{(\varphi)},
\mathcal{E}_{(A)}^\nu, \mathcal{E}_{(B)}^{\mu\nu}\big)$, defined by
\bea
\mathcal{E}^{(R)}_{\mu\nu} &=&\mathcal{G}_{\mu\nu}+T^{(\varphi)}_{\mu\nu}
+T^{(A)}_{\mu\nu}+T^{(B)}_{\mu\nu} \, , \nn \\
\mathcal{E}_{(\varphi)}&=&\frac{d\chi}{d\varphi}\nabla^\mu \varphi\nabla_\mu \varphi
-2\nabla_\mu (\chi\nabla^\mu\varphi)+\frac{dV}{d\varphi}+\frac{d\lambda}{d\varphi} R
+\frac{dN}{d\varphi}\frac{L_{(A)}}{N}+\frac{dP}{d\varphi}\frac{L_{(B)}}{P} \, , \nn \\
\mathcal{E}_{(A)}^\nu &=& 2\alpha A_\mu\mathcal{E}_{(B)}^{\mu\nu}
-12\alpha PH^{\nu\alpha\beta}F_{\alpha\beta}-4\nabla_\mu (NF^{\mu\nu}) \, , \nn \\
\mathcal{E}_{(B)}^{\mu\nu} &=& -6\nabla_\rho (PH^{\rho\mu\nu}) \, , \label{EoEoMgel}
\eea
where
\bea
\mathcal{G}_{\mu\nu}&=&\lambda \Big(R_{\mu\nu}-\frac{1}{2}g_{\mu\nu}R\Big)
+g_{\mu\nu}\nabla^\rho\nabla_\rho\lambda-\nabla_\mu\nabla_\nu\lambda \, , \nn \\
T^{(\varphi)}_{\mu\nu}&=&\chi\nabla_\mu \varphi\nabla_\nu \varphi
-\frac{1}{2}g_{\mu\nu}L_{(\varphi)} \, , \nn \\
T^{(A)}_{\mu\nu}&=& 2NF_{\mu\sigma}F_\nu^{~\sigma}-\frac{1}{2}g_{\mu\nu}L_{(A)} \, ,\nn \\
T^{(B)}_{\mu\nu}&=& 3PH_{\mu\alpha\beta}H_\nu^{~\alpha\beta}-\frac{1}{2}g_{\mu\nu}L_{(B)}
\, , \label{EqforGrav}
\eea
the variation of the Lagrangian (\ref{GenLagran}) with respect to the metric tensor and
matter fields $\varphi^{(i)}$ yields
\be
\delta\mathcal{L} =\sqrt{-g}\big(\mathcal{E}^{(R)}_{\mu\nu}\delta g^{\mu\nu}
+\mathcal{E}_{(\varphi)}\delta\varphi
+\mathcal{E}_{(A)}^\nu\delta A_\nu
+\mathcal{E}_{(B)}^{\mu\nu}\delta B_{\mu\nu}\big)
+\sqrt{-g}\nabla_\mu \Theta^\mu\big(\varphi^{(i)};\delta\varphi^{(i)}\big) \, , \label{VarLagran}
\ee
where the boundary term $\Theta^\mu\big(\varphi^{(i)};\delta\varphi^{(i)}\big)$, locally constructed
out of the fields $\varphi^{(i)}$, $\delta\varphi^{(i)}$ and their derivatives,
is given by
\bea
\Theta^\mu&=&\Theta_{(R)}^\mu(\delta g)
+\Theta_{(\varphi)}^\mu(\delta \varphi)+\Theta_{(A)}^\mu(\delta A)
+\Theta_{(B)}^\mu(\delta B) \, , \nn \\
\Theta_{(R)}^\mu&=& 2\lambda g^{\rho[\mu}\nabla^{\sigma]}\delta g_{\rho\sigma}
+\delta g^{\mu\sigma}\nabla_\sigma \lambda
+g^{\alpha\beta}\delta g_{\alpha\beta}\nabla^\mu \lambda \, , \nn \\
\Theta_{(\varphi)}^\mu&=& 2\chi \delta\varphi\nabla^\mu \varphi \, ,  \nn \\
\Theta_{(A)}^\mu&=&-(12\alpha PA_\rho H^{\rho\mu\nu}-4NF^{\mu\nu})\delta A_\nu
\, ,  \quad
\Theta_{(B)}^\mu=6PH^{\mu\alpha\beta}\delta B_{\alpha\beta} \, .
\label{BoundTerm}
\eea

Let $\zeta^\mu$ denote any smooth vector field on the spacetime. Under the coordinate
transformation $x^\mu\rightarrow x^\mu-\zeta^\mu$, the variation of the fields
$\varphi^{(i)}$ behaves as their Lie derivative along the vector field $\zeta^\mu$,
i.e., $\delta\varphi^{(i)}\rightarrow \mathcal{L}_\zeta \varphi^{(i)}$. This makes it
possible to reexpress the part in the bracket of Eq. (\ref{VarLagran}) as the divergence
of a vector field \cite{CiteHJPY,IWaldC,GaoWaldC}, namely,
\be
\mathcal{E}_{(i)}\mathcal{L}_\zeta \varphi^{(i)}
=-2\nabla_\mu(\mathcal{E}^{\mu\nu}\zeta_\nu) \, , \quad
\mathcal{E}^{\mu\nu}= \mathcal{E}_{(R)}^{\mu\nu}
-\mathcal{E}_{(B)}^{\mu\sigma}B^\nu_{~\sigma}-\frac{1}{2}\mathcal{E}_{(A)}^{\mu}A^\nu
\, , \label{Emunu}
\ee
where the tensor $\mathcal{E}^{\mu\nu}$ is a linear combination of the expressions
for the Euler-Lagrange equations of motion $\mathcal{E}_{(R)}^{\mu\nu}$, $\mathcal{E}_{(A)}^\nu$ and
$\mathcal{E}_{(B)}^{\mu\nu}$. Unlike the case of pure gravity theories, $\mathcal{E}^{\mu\nu}$
does not have to be symmetric.
Replacing the variation by the Lie derivative along the vector $\zeta^\mu$ in
Eq. (\ref{VarLagran}), one can follow \cite{CiteHJPY,TPadman} to define an off-shell
Noether current $J^\mu$ that includes contributions from matter fields as
\bea
J^\mu&=&2\mathcal{E}^{\mu\nu}\zeta_\nu+\zeta^\mu L
-\Theta^\mu\big(\varphi^{(i)};\mathcal{L}_\zeta\varphi^{(i)}\big)
=J_{(R)}^\mu+J_{(\varphi)}^\mu+J_{(A)}^\mu+J_{(B)}^\mu  \nn \\
&=&\nabla_\nu K^{\mu\nu}\, . \label{NoCurrGen}
\eea
In the above equation, all the off-shell currents $J_{(n)}^\mu$, here and
in what follows $n=(R,\varphi,A,B)$, take the forms
\bea
J_{(R)}^\mu&=&2\mathcal{G}^{\mu\nu}\zeta_\nu+\zeta^\mu L_{(R)}
-\Theta_{(R)}^\mu\big(\mathcal{L}_\zeta g\big)
=\nabla_\nu K_{(R)}^{\mu\nu} \, , \nn \\
J_{(\varphi)}^\mu&=&2T_{(\varphi)}^{\mu\nu}\zeta_\nu+\zeta^\mu L_{(\varphi)}
-\Theta_{(\varphi)}^\mu\big(\mathcal{L}_\zeta \varphi\big)=0  \, , \nn \\
J_{(A)}^\mu&=&2T_{(A)}^{\mu\nu}\zeta_\nu+\zeta^\mu L_{(A)}
-\Theta_{(A)}^\mu\big(\mathcal{L}_\zeta A\big)
-\mathcal{E}_{(A)}^{\mu}A^\nu\zeta_\nu
-18\alpha PH^{\mu\alpha\beta}A_{[\nu}F_{\alpha\beta]}\zeta^\nu \nn \\
&=&\nabla_\nu K_{(A)}^{\mu\nu}\, , \nn \\
J_{(B)}^\mu&=&2T_{(B)}^{\mu\nu}\zeta_\nu+\zeta^\mu L_{(B)}
-\Theta_{(B)}^\mu\big(\mathcal{L}_\zeta B\big)
-2\mathcal{E}_{(B)}^{\mu\sigma}B^\nu_{~\sigma}\zeta_\nu
+18\alpha PH^{\mu\alpha\beta}A_{[\nu}F_{\alpha\beta]}\zeta^\nu  \nn \\
&=&\nabla_\nu K_{(B)}^{\mu\nu}\, , \label{CurrRvAB}
\eea
while the off-shell Noether potentials $K^{\mu\nu}$ and $K_{(n)}^{\mu\nu}$ are given by
\bea
K^{\mu\nu}&=&K_{(R)}^{\mu\nu}+K_{(\varphi)}^{\mu\nu}+K_{(A)}^{\mu\nu}+K_{(B)}^{\mu\nu} \, , \nn \\
K_{(R)}^{\mu\nu}&=&2\lambda\nabla^{[\mu}\zeta^{\nu]}
+4\zeta^{[\mu}\nabla^{\nu]}\lambda \, , \quad
K_{(\varphi)}^{\mu\nu}=0\, , \nn \\
K_{(A)}^{\mu\nu}&=&(12\alpha PA_\rho H^{\rho\mu\nu}-4NF^{\mu\nu})A_\sigma\zeta^\sigma
\, , \nn \\
K_{(B)}^{\mu\nu}&=& 12PH^{\mu\nu\rho}B_{\rho\sigma}\zeta^\sigma
\, . \label{NoePoten}
\eea
Note that the off-shell Noether potential $K_{(B)}^{\mu\nu}$ in Eq. (\ref{NoePoten}) is consistent
with the potential for two-form field got through the covariant phase space approach in \cite{Rogatko}
when $\zeta^\mu$ is a Killing vector and $P=-exp(-\alpha\phi)/12$.

Assume now that the smooth vector field $\zeta^\mu$ respects the symmetry of spacetime
by treating it as a Killing vector $\xi^\mu$. We follow \cite{CiteHJPY} to introduce the off-shell
ADT current $J_{ADT}^\mu$ associated with the Killing vector $\xi^\mu$ by
\bea
J_{ADT}^\mu &=&\delta\mathcal{E}^{\mu\nu}\xi_\nu
+\frac{1}{2}g^{\alpha\beta}\delta g_{\alpha\beta}\mathcal{E}^{\mu\nu}\xi_\nu
+\mathcal{E}^{\mu\nu}\delta g_{\nu\sigma}\xi^\sigma
+\frac{1}{2}\xi^\mu\mathcal{E}_{(i)}\delta \varphi^{(i)}\nn \\
&=&\nabla_\nu Q_{ADT}^{\mu\nu}\, , \label{ADTcurr}
\eea
where $Q_{ADT}^{\mu\nu}$ is just the off-shell ADT potential corresponding to
the ADT current. In terms of the variation of the Lagrangian (\ref{GenLagran}) and
the definition of the off-shell Noether current $J^\mu$, the ADT potential which
is in one-to-one correspondence with the off-shell Noether potential
can be presented by
\bea
Q_{ADT}^{\mu\nu}&=&\frac{1}{2}\frac{1}{\sqrt{-g}}\delta\big(\sqrt{-g}K^{\mu\nu}\big)
-\xi^{[\mu}\Theta^{\nu]}\big(\varphi^{(i)};\delta\varphi^{(i)}\big)  \nn \\
&=&Q_{(R)}^{\mu\nu}+Q_{(\varphi)}^{\mu\nu}+Q_{(A)}^{\mu\nu}
+Q_{(B)}^{\mu\nu}\, , \nn \\
Q_{(n)}^{\mu\nu}&=&\frac{1}{2}\frac{1}{\sqrt{-g}}\delta\big(\sqrt{-g}K_{(n)}^{\mu\nu}\big)
-\xi^{[\mu}\Theta_{(n)}^{\nu]} \, . \label{ADTpotenQ}
\eea
In Eq. (\ref{ADTpotenQ}), the Killing vector $\xi^\mu$ is treated as a fixed background, namely,
$\delta\xi^\mu=0$, and the vector field $\zeta^\mu$ in $K_{(n)}^{\mu\nu}$ given by
Eq. (\ref{NoePoten}) is substituted by the Killing vector $\xi^\mu$.
The quantities $Q_{(R)}^{\mu\nu}$, $Q_{(\varphi)}^{\mu\nu}$, $Q_{(A)}^{\mu\nu}$ and
$Q_{(B)}^{\mu\nu}$ denote the contributions from the gravitational field, scalar field,
U(1) gauge field and two-form field respectively. The potential $Q_{(R)}^{\mu\nu}$ can
be seen as a special case of the off-shell ADT potential in the work \cite{JJPengHornCC}.
It is worth noting that $Q_{(\varphi)}^{\mu\nu}$ and $Q_{(B)}^{\mu\nu}$ are equivalent to those obtained
via BBC method in \cite{BarnichSca,ComMuNI,ComperH}. What is more, if we directly adopt the BBC method
\cite{BarnichB,Barnich,BarnichC,BCintegC} to compute the superpotential for the Lagrangian
(\ref{GenLagran}), we shall get the result that is equivalent to the off-shell ADT potential
$Q_{ADT}^{\mu\nu}$. This implies that both the off-shell generalized ADT method
\cite{KimKY,CiteHJPY} and BBC method are equivalent for the Lagrangian (\ref{GenLagran}).
In other words, the off-shell generalized ADT method provides another way to derive the
superpotential of the Lagrangian (\ref{GenLagran}) in the BBC method. Besides, it was
shown that the off-shell ADT potential is completely equivalent to that got through the
covariant phase space approach in the presence of generic matter fields in \cite{CiteHJPY}.

By following the BBC approach \cite{BarnichB,Barnich,BarnichC,BCintegC} to incorporate
a single parameter path characterized by a parameter $s$, $s\in[0,1]$, in the space
of solutions, we define the covariant formulation of conserved charges associated with the
Noether potential $Q_{ADT}^{\mu\nu}$ in Eq. (\ref{ADTpotenQ}) by \cite{KimKY,CiteHJPY}
\be
\mathcal{Q}=\frac{1}{8\pi}\int_0^1 ds \int d\Sigma_{\mu\nu} Q_{ADT}^{\mu\nu}
\big(\varphi^{(i)};s\big)
\, , \label{QdefineAn}
\ee
where $d\Sigma_{\mu\nu}=\frac{1}{2}\frac{1}{(D-2)!}
\epsilon_{\mu\nu\mu_1\mu_2\cdot\cdot\cdot\mu_{(D-2)}}dx^{\mu_1}\wedge\cdot\cdot\cdot
\wedge dx^{\mu_{(D-2)}}$ with $\epsilon_{012\cdot\cdot\cdot(D-1)}=\sqrt{-g}$.
Eq. (\ref{QdefineAn}) can be a proposal of the formalism for
the conserved charge, defined in the interior region or at the asymptotical infinity,
for any covariant gravity theory with the Lagrangian (\ref{GenLagran}) whenever its
integration is well-defined \cite{BCintegC}. Since the Lagrangian (\ref{GenLagran})
also includes the scalar-tensor theories of gravity, Einstein gravity coupled to a
scalar field non-minimally and Einstein-Maxwell-Dilaton gravity
as its special cases, formalism (\ref{QdefineAn}) is applicable to these gravity theories.
To see this, we shall explicitly compute the energy of the static charged black holes
in Einstein-Maxwell-Dilaton gravity in the \ref{appA}. It will be shown that the
conserved charges defined through Eq. (\ref{QdefineAn}) are conformally invariant
in the \ref{appB}.

\section{Mass and angular momentum of Kerr-Sen black hole in the string frame}\label{sectwo}

In this section, we will employ the covariant formalism (\ref{QdefineAn}) to compute
the mass and angular momentum of the well-known Kerr-Sen black hole \cite{KerrSen}
in the string frame. The effect from the scalar field, U(1) gauge field and two-form
field will be fully considered.
The Kerr-Sen black hole is a four-dimensional solution in the
low-energy effective field theory describing heterotic string theory. In the string
frame, the effective action reads
\be
\mathcal{L}= \sqrt{-g}e^{-\varphi}\Big(R
+\nabla^\mu \varphi\nabla_\mu \varphi
-\frac{1}{8}F^{\mu\nu}F_{\mu\nu}
-\frac{1}{12}H^{\rho\mu\nu}H_{\rho\mu\nu}\Big)
\, .   \label{KSLagran}
\ee
Comparing the above equation with the general Lagrangian (\ref{GenLagran}),
we have
\be
\lambda=\chi=-8N=-12P=e^{-\varphi} \, , \quad V=0 \, , \quad
\alpha=\frac{1}{4} \, .
\ee
In Boyer-Lindquist coordinate, the Kerr-Sen black hole in the string frame takes
the form
\bea
\frac{ds^2}{\Sigma}&=& -\frac{\Delta}{(\Sigma+2ms^2r)^2}(dt-a\sin^2\theta d\phi)^2
+\Big(\frac{dr^2}{\Delta}+d\theta^2\Big) \nn \\
&&+\frac{\sin^2\theta}{(\Sigma+2ms^2r)^2}[adt-(\Delta+2mc^2r)d\phi]^2
\, , \nn \\
e^{-\varphi}&=&1+\frac{2ms^2r}{\Sigma}\, ,\quad s=\sinh\gamma \, ,
\quad c=\cosh\gamma\, , \nn \\
A&=&\frac{4mcsr}{\Sigma+2ms^2r}(dt-a\sin^2\theta d\phi) \, ,  \nn \\
B_{\mu\nu}&=&\frac{4ams^2r\sin^2\theta}{\Sigma+2ms^2r}\delta^{[t}_\mu\delta^{r]}_\nu
 \, , \label{KerrSmetric}
\eea
where the parameters $m$, $a$, $\gamma$ are integration constants associated with
the mass, angular momentum and electric charge, while the functions $\Delta$ and $\Sigma$
are given by
\be
\Delta=r^2+a^2-2mr\, , \quad \Sigma=r^2+a^2\cos^2\theta \, . \label{DeltaSig}
\ee

We now calculate the mass of the Kerr-Sen black hole. The related timelike Killing vector
$\xi^\mu=(-1,0,0,0)$. The infinitesimal variation of the fields is determined by letting the
constants $(m,a,c,s)$ change as
\be
m\rightarrow m+dm \, , \quad a\rightarrow a+da \, , \quad
c\rightarrow c+dc \, , \quad s\rightarrow s+ds \, . \label{KScosntch}
\ee
On basis of the formulation of the off-shell ADT potential in Eq. (\ref{ADTpotenQ}),
the $(t,r)$ components of the quantities $Q_{(n)}^{\mu\nu}$ are presented by
\bea
\sqrt{-g}Q_{(R)}^{tr}&=&2\sin\theta[d(m)+2msd(s)+s^2d(m)]+\mathcal{O}\Big(\frac{1}{r}\Big) \, , \nn \\
\sqrt{-g}Q_{(\varphi)}^{tr}&=&\mathcal{O}\Big(\frac{1}{r^3}\Big) \, , \quad
\sqrt{-g}Q_{(A)}^{tr}=\mathcal{O}\Big(\frac{1}{r^3}\Big) \, ,\quad
\sqrt{-g}Q_{(B)}^{tr}=\mathcal{O}\Big(\frac{1}{r^5}\Big) \, . \label{KSQtr}
\eea
The above equation shows that the contributions to the ADT potential from the matter
fields are not totally zero. However, since $\sqrt{-g}Q_{(\varphi)}^{tr}$,
$\sqrt{-g}Q_{(A)}^{tr}$ and $\sqrt{-g}Q_{(B)}^{tr}$ fall off very fast at infinity,
the matter fields finally make no contribution to the mass of the Kerr-Sen black hole.
In terms of the formalism (\ref{QdefineAn}) of the conserved charge, we integrate
$\sqrt{-g}Q^{tr}=\sqrt{-g}(Q_{(R)}^{tr}+Q_{(\varphi)}^{tr}+Q_{(A)}^{tr}+Q_{(B)}^{tr})$
on the surface at infinity and get the mass
\be
M_{KS}=m(1+s^2) \, ,
\ee
which agrees with the Brown-York quasilocal energy \cite{JingHuang,BoseNai} calculated
in the Einstein frame.

The process to obtain the angular momentum of the Kerr-Sen black hole is parallel to the
mass. Substituting the Killing vector $\xi^\mu=(0,0,0,1)$ into Eq. (\ref{ADTpotenQ}), we
get the $(t,r)$ component of the ADT potential
\be
\sqrt{-g}\tilde{Q}^{tr}=3\sin^3\theta
[(1+s^2)d(ma)+mad(s^2)]+\mathcal{O}\Big(\frac{1}{r}\Big) \, ,
\ee
which is associated with the angular momentum.
Integration of $\sqrt{-g}\tilde{Q}^{tr}$ on the surface at infinity further yields
the angular momentum
\be
J_{KS}=ma(1+s^2) \, .
\ee
Both the mass $M_{KS}$ and the angular momentum $J_{KS}$ satisfy the first law of
black hole thermodynamics and they are consistent with the ones via the ADM
formalism in the Einstien frame.

\section{Mass and angular momentum of the five-dimensional black string}\label{secthree}

In this section, the covariant formalism (\ref{QdefineAn}) is applied to compute
the mass and angular momentum of the five-dimensional rotating charged black
string constructed by Mahapatra in \cite{5Dblackst}. This black string characterized by
four parameters $(m,a,\gamma_1,\gamma_2)$ is an exact solution for the Lagrangian
(\ref{KSLagran}). The metric of the black string admits a generalized Killing-Yano
tensor \cite{KillYano}. In the string frame, it is presented by
\bea
ds^2&=&\frac{-\Sigma(\Sigma-2mr)}{\Xi^2}
\Big(dt+\frac{mar(1+c_1c_2)\sin^2\theta}{\Sigma-2mr}d\phi\Big)^2
+\Sigma\Big(\frac{dr^2}{\Delta}+d\theta^2\Big) \nn \\
&&+\frac{\Sigma\Delta\sin^2\theta}{\Sigma-2mr}d\phi^2
+\Big(dz+\frac{c_1s_2}{2s_1}A\Big)^2 \, ,\nn \\
s_i&=&\sinh\gamma_i \, ,
\quad c_i=\cosh\gamma_i \, , \quad i=1,2 \, , \label{Blackstr}
\eea
where $\Xi=\Sigma-m(1-c_1c_2)r$, and both the functions $\Sigma$ and $\Delta$
have been given in Eq. (\ref{DeltaSig}). The matter fields take the forms
\bea
e^{-\varphi}&=&\frac{\Xi}{\Sigma} \, , \nn \\
A&=&\frac{2ms_1r}{\Xi}(dt-a\sin^2\theta d\phi) \, , \nn \\
B&=&\frac{1}{2s_1}[(1-c_1c_2)dt-c_1s_2dz]\wedge A \, . \label{Bstrimfields}
\eea

To compute the mass and angular momentum per unit length of the black string,
we take an infinitesimal parametrization of a single parameter path by letting
the integration constants $(m, a,c_i,s_i)$ fluctuate as
\be
m\rightarrow m+dm \, , \quad a\rightarrow a+da \, , \quad
c_i\rightarrow c_i+dc_i \, , \quad
s_i\rightarrow s_i+ds_i \, . \label{Bstrcosntch}
\ee
Combined with the timelike Killing vector $\xi^\mu=-\delta_t^\mu$, which is the symmetry
corresponding to the mass, the $(t,r)$ component of the ADT potential reads
\bea
\sqrt{-g}Q^{tr}&=&\sqrt{-g}(Q_{(R)}^{tr}+Q_{(\varphi)}^{tr}+Q_{(A)}^{tr}+Q_{(B)}^{tr}) \, , \nn \\
\sqrt{-g}Q_{(R)}^{tr}&=&\sin\theta[(1+c_1c_2)d(m)+md(c_1c_2)]
+\mathcal{O}\Big(\frac{1}{r}\Big) \, , \nn \\
\sqrt{-g}Q_{(\varphi)}^{tr}&=&\mathcal{O}\Big(\frac{1}{r^3}\Big) \, , \quad
\sqrt{-g}Q_{(A)}^{tr}=\mathcal{O}\Big(\frac{1}{r^3}\Big) \, ,\quad
\sqrt{-g}Q_{(B)}^{tr}=\mathcal{O}\Big(\frac{1}{r^3}\Big) \, . \label{BStrQtr}
\eea
Thus the mass per unit length of the black string is read off as
\be
M_{bstr}=\frac{1}{2}m(1+c_1c_2) \, . \label{massbstr}
\ee
By assuming that the Kiling vector $\xi^\mu=\delta_\phi^\mu$, the angular momentum per
unit length of the black string can be got by performing an analogous calculation. We
present it by
\be
J_{bstr}=\frac{1}{2}ma(1+c_1c_2)=M_{bstr}a \, . \label{anmbstr}
\ee
Both the mass and angular momentum per unit length coincide with the ones via
the ADM formalism.

\section{Summary} \label{secfour}

In the present work, we have investigated the conserved charges in the low-energy
effective field theory of heterotic string theory by building a one-to-one
correspondence between the ADT potential and the off-shell Noether potential
like in \cite{KimKY,CiteHJPY}.
Starting with the general Lagrangian (\ref{GenLagran}), which also includes
the scalar-tensor theories of gravity, Einstein gravity coupled to a scalar
field non-minimally and Einstein-Maxwell-Dilaton gravity as its special cases,
we derive the off-shell Noether currents and potentails in Eqs. (\ref{CurrRvAB})
and (\ref{NoePoten}). Establishing the relationship between the ADT potential
and the off-shell Noether potential, the covariant formalism (\ref{QdefineAn})
for the conserved charge is defined in terms of the ADT potential in
Eq. (\ref{ADTpotenQ}). Then the covariant formalism (\ref{QdefineAn}) is applied
to compute the mass and angular momentum of the four-dimensional Kerr-Sen black hole
(\ref{KerrSmetric}) and the five-dimensional rotating charged black string
(\ref{Blackstr}) in the string frame. These results coincide with those via ADM
formalism in the Einstein frame and they satisfy the first law of thermodynamics.
In \ref{appA}, the covariant formalism (\ref{QdefineAn}) is also used
to calculate the energy of the general static charged black holes in arbitrary
dimensions within the framework of Einstein-Maxwell-Dilaton gravity. In the
\ref{appB}, we have proved that the conserved charges are invariant under
conformal transformation.

Although we only take into account of the mass and angular momentum of the Kerr-Sen
black hole and the black string, one can expect that the formalism (\ref{QdefineAn})
for conserved charges is also applicable to the higher-dimensional rotating charged
black holes with NUT parameters in \cite{Chow} and the five-dimensional black rings in
\cite{Yazadjiev,Elvang}. A future interesting application of the off-shell Noether currents
and potentails in Eqs. (\ref{CurrRvAB}) and (\ref{NoePoten}) is to adopt the
method in \cite{{MaPadma1,MaPadma2}} to investigate the statistical entropy of
black holes and black rings in the low-energy limit of heterotic string theory by
including the contributions from matter fields.

\section*{Acknowledgments}

We would like to thank the anonymous referees for their valuable suggestions and comments.
This work was supported by the Natural Science Foundation of China under Grant
No. 11275157 and No. 11505036. It was also partially supported by the Doctoral
Research Fund of Guizhou Normal University in 2014 and Guizhou province science
and technology innovation talent team [Grant No. (2015)4015].

\appendix
\section{Energy of static charged black holes in Einstein-Maxwell-Dilaton gravity}\label{appA}

In this appendix, we take the theory of Einstein-Maxwell-Dilaton gravity as a
special case of the general Lagrangian (\ref{GenLagran}) and make use of the
covariant formalism (\ref{QdefineAn}) to calculate the energy of static
and charged black holes in such a theory. Till now, much interest has been
attracted by seeking exact solutions in the theory
\cite{Sheykhi,SheyDHen,ChaGoSo,CaiJS,CHMann,SQWu,KunzML}. A lot of static black hole
solutions with various asymptotic structures have been constructed. We aim to
provide another novel method to compute the conserved charges of these solutions. The
Lagrangian of the theory of Einstein-Maxwell-Dilaton gravity reads
\be
\mathcal{L}=\sqrt{-g}\Big(R-\frac{1}{2}\nabla^\mu \varphi\nabla_\mu \varphi+V(\varphi)
+N(\varphi)F^{\mu\nu}F_{\mu\nu}\Big)
\, . \label{EMDLagran}
\ee
The metric and U(1) gauge field for the static and charged black
holes in $d$ dimensions are assumed to take the general form
\bea
ds^2&=&-W(r)^2dt^2+\frac{dr^2}{U(r)^2}+r^2h_{ij}dx^idx^j \, , \nn \\
A&=& A_t(r) dt+A_i(r,x^j)dx^i
\, , \label{SpBHinEMD}
\eea
where $i,j=1,\cdot\cdot\cdot,(d-2)$ and $h_{ij}$ is a function of coordinates
$x^i$, which span a $(d-2)$-dimensional hypersurface.

The Killing vector associated with the energy is chosen as $\xi^\mu=-\delta_t^\mu$,
and $h_{ij}$ is fixed, namely, $\delta h_{ij}=0$.
By computing the ADT potential in terms of Eq. (\ref{ADTpotenQ}), we have
\bea
\sqrt{-g}Q^{tr}&=&\sqrt{-g}(Q_{(R)}^{tr}+Q_{(\varphi)}^{tr}+Q_{(A)}^{tr}) \, , \nn \\
\sqrt{-g}Q_{(R)}^{tr}&=&\frac{1}{2}\sqrt{-g}g^{r\mu}g^{i\nu}\nabla_i\delta g_{\mu\nu}
=-(d-2)\sqrt{h}r^{d-3}W\delta U\, , \nn \\
\sqrt{-g}Q_{(\varphi)}^{tr}&=&-\frac{1}{2}\sqrt{h}r^{d-2}WU\partial_r\varphi\delta\varphi \, , \nn \\
\sqrt{-g}Q_{(A)}^{tr}&=&2A_t\delta(\sqrt{-g}NF^{tr})
+2\sqrt{-g}NF^{rj}\delta A_j \, . \label{EMDQtr}
\eea
The U(1) gauge field $A_\mu$ satisfies the equation of motion $-4\nabla_\mu(NF^{\mu\nu})=0$.
Its $t$ component reads $\partial_r(\sqrt{-g}NF^{tr})=0$, which leads to that $\sqrt{-g}NF^{tr}$ only
depends on the coordinates $x^i$. This implies that the fall-off condition of the gauge field at
infinity completely determines whether the first term in $\sqrt{-g}Q_{(A)}^{tr}$ makes contribution
to the total energy or not. In general
cases, $A_t\sim \mathcal{O}(1/r)$, which means that the contribution from the first term in
$\sqrt{-g}Q_{(A)}^{tr}$ usually can be neglected. Integration of the ADT potential
\be
\sqrt{-g}Q^{tr}=-\frac{1}{2}\sqrt{h}r^{d-4}W[2(d-2)r\delta U
+r^2U\partial_r\varphi\delta\varphi
-4NUh^{ij}\partial_r A_i\delta A_j]
\ee
over the surface at infinity in terms of the formalism (\ref{QdefineAn}) yields the energy
of the static charged black holes if the integration is well-defined \cite{BCintegC}. It is
very interesting to apply Eq. (\ref{EMDQtr}) to investigate the first law of thermodynamics
of the static black holes in
\cite{Sheykhi,SheyDHen,ChaGoSo,CaiJS,CHMann} by considering the contribution from the scalar
field along the line of works \cite{HenMTZ,LuPaPo,LiuLu}.

\section{Conformal property of the conserved charge}\label{appB}

In the present appendix, we investigate conformal property of the conserved
charge in the low-energy effective field theory describing heterotic string theory.
Performing the conformal transformation
\be
g_{\mu\nu}=\Omega^2(\varphi)\hat{g}_{\mu\nu} \, ,\qquad
\Omega=\lambda^{-\frac{1}{D-2}} \, \label{ConfTrg}
\ee
to the Lagrangian (\ref{GenLagran}), we obtain the Lagrangian in the Einstein frame
\bea
\hat{\mathcal{L}}&=&\sqrt{-\hat{g}}(\hat{L}_{(R)}+\hat{L}_{(\varphi)}
+\hat{L}_{(A)}+\hat{L}_{(B)}) +\hat{\mathcal{L}}_{Bound}\nn \, , \\
\hat{L}_{(R)} &=&\hat{R}(\hat{g}) \, ,
\quad \hat{L}_{(\varphi)}
=\Big[\Omega^{D-2}\chi-(D-1)(D-2)\Big(\frac{\partial\ln\Omega}{\partial\varphi}\Big)^2\Big]
\hat{\nabla}^\mu \varphi\hat{\nabla}_\mu \varphi
+\Omega^D V  \nn \, , \\
\hat{L}_{(A)} &=& \Omega^{D-4}N\hat{F}^{\mu\nu}\hat{F}_{\mu\nu}  \, , \quad
\hat{L}_{(B)} = \Omega^{D-6}P\hat{H}^{\rho\mu\nu}\hat{H}_{\rho\mu\nu}\, .
\label{ConfTraction}
\eea
In the above equation and what follows, $\hat{F}_{\mu\nu}=F_{\mu\nu}$,
$\hat{H}_{\rho\mu\nu}=H_{\rho\mu\nu}$ and all the quantities with
`` $^{\hat{}}$ " correspond to the ones in the Einstein frame, whose
indices are raised or lowered by the metric tensor $\hat{g}^{\mu\nu}$
or $\hat{g}_{\mu\nu}$. The total divergence term
$\hat{\mathcal{L}}_{Bound}=-2(D-1)\sqrt{-\hat{g}}\hat{\Box}\ln\Omega$
makes no contribution to the field equations and the conserved
charges.

The surface terms got from the variation of the Lagrangian (\ref{ConfTraction})
are read off as
\bea
\hat{\Theta}^\mu&=&\hat{\Theta}_{(R)}^\mu+\hat{\Theta}_{(\varphi)}^\mu
+\hat{\Theta}_{(A)}^\mu+\hat{\Theta}_{(B)}^\mu \, , \nn \\
\hat{\Theta}_{(R)}^\mu&=&
\frac{(D-1)}{\sqrt{-\hat{g}}}\delta\big(\sqrt{-\hat{g}}\hat{\nabla}^\mu\ln\Omega^2\big)
+\frac{(D-1)(D-2)}{2}(\delta\ln\Omega^2)\hat{\nabla}^\mu\ln\Omega^2  \nn \\
&&+\Omega^{D}\Theta_{(R)}^\mu
\, , \nn \\
\hat{\Theta}_{(\varphi)}^\mu&=& \Omega^{D}\Theta_{(\varphi)}^\mu
-\frac{(D-1)(D-2)}{2}(\delta\ln\Omega^2)\hat{\nabla}^\mu\ln\Omega^2\, ,  \nn \\
\hat{\Theta}_{(A)}^\mu&=&\Omega^{D}\Theta_{(A)}^\mu
\, ,  \quad
\hat{\Theta}_{(B)}^\mu=\Omega^{D}\Theta_{(B)}^\mu \, .\label{SurTCFaction}
\eea
With help of Eq. (\ref{NoePoten}), the off-shell Noether potentials
$\hat{K}^{\mu\nu}$ and $\hat{K}_{(n)}^{\mu\nu}$ in the Einstein frame are given
by
\bea
\hat{K}^{\mu\nu}&=&\hat{K}_{(R)}^{\mu\nu}+\hat{K}_{(\varphi)}^{\mu\nu}
+\hat{K}_{(A)}^{\mu\nu}+\hat{K}_{(B)}^{\mu\nu} \, , \nn \\
\hat{K}_{(R)}^{\mu\nu}&=&\Omega^{D}K_{(R)}^{\mu\nu}
+2(D-1)\zeta^{[\mu}\hat{\nabla}^{\nu]}\ln\Omega^2\, , \nn \\
\hat{K}_{(\varphi)}^{\mu\nu}&=&0\, ,\qquad
\hat{K}_{(A)}^{\mu\nu}=\Omega^{D}K_{(A)}^{\mu\nu}
\, , \qquad
\hat{K}_{(B)}^{\mu\nu}=\Omega^{D}K_{(B)}^{\mu\nu}
\, . \label{CFinNoePoten}
\eea
Here we have assumed that the vector field $\zeta^\mu$ is unchanged under the
conformal transformation. In terms of the above surface terms and the Noether
potentials, we get the ADT potential corresponding to the Lagrangian
(\ref{ConfTraction}), which takes the form
\be
\sqrt{-\hat{g}}\hat{Q}_{ADT}^{\mu\nu}=\sqrt{-g}Q_{ADT}^{\mu\nu}
\, . \label{ADTpotenEinF}
\ee
Eq. (\ref{ADTpotenEinF}) indicates that the conserved charges in the context of
the low-energy effective field theory describing heterotic string theory
are invariant under conformal transformation. Up to this point, the off-shell
generalized ADT formalism is different from the original ADT formalism. Conserved
charges defined through the latter depend on the asymptotic behavior of the
conformal factor under conformal transformation and they are conformally invariant
as long as the conformal factor goes to unity at infinity \cite{ConfTofDT}.

As a result, it is fully feasible to adopt the formalism (\ref{QdefineAn}) to
compute the conserved charges of the Kerr-Sen black hole and the five-dimensional
black string in the Einstein frame. To do this, we need reexpress the string-frame
metric $g_{\mu\nu}$ as the Einstein-frame metric $\hat{g}_{\mu\nu}$ by letting
$g_{\mu\nu}=e^{\varphi}\hat{g}_{\mu\nu}$. The Lagrangian (\ref{KSLagran})
accordingly becomes
\be
\mathcal{\hat{L}}= \sqrt{-\hat{g}}\Big(\hat{R}(\hat{g})
-\frac{1}{2}\hat{\nabla}^\mu \varphi\hat{\nabla}_\mu \varphi
-\frac{1}{8}e^{-\varphi}\hat{F}^{\mu\nu}\hat{F}_{\mu\nu}
-\frac{1}{12}e^{-2\varphi}\hat{H}^{\rho\mu\nu}\hat{H}_{\rho\mu\nu}\Big)
\,  \label{ConfTrKSaction}
\ee
Performing analogous calculations as in the string frame on a basis of the
above Lagrangian, we obtain the mass and angular momentum that are completely
consistent with those corresponding to the Lagrangian (\ref{KSLagran}).


\begin{thebibliography}{99}

\bibitem{GiMasol}
G.W. Gibbons and K. Maeda, \emph{Nucl. Phys. B} \textbf{298}, 741 (1988).

\bibitem{GaHoStsol}
D. Garfinkle, G. T. Horowitz and A. Strominger,
\emph{Phys. Rev. D} \textbf{43}, 3140 (1991).

\bibitem{SurDasS}
S. Sur, S. Das and S. SenGupta,
\emph{JHEP} \textbf{0510}, 064 (2005),
arXiv:hep-th/0508150.

\bibitem{KerrSen}
A. Sen, \emph{Phys. Rev. Lett.} \textbf{69}, 1006 (1992),
arXiv:hep-th/9204046.

\bibitem{Chow}
D.D.K. Chow, \emph{Class. Quantum Grav.} \textbf{27}, 205009 (2010),
arXiv:0811.1264 [hep-th].

\bibitem{5Dblackst}
S. Mahapatra, \emph{Phys. Rev. D} \textbf{50}, 947 (1994),
arXiv:hep-th/9301125.

\bibitem{Yazadjiev}
S.S. Yazadjiev, \emph{Phys. Rev. D}  \textbf{73}, 124032 (2006), arXiv:hep-th/0512229.

\bibitem{Elvang}
H. Elvang, \emph{Phys. Rev. D} \textbf{68}, 124016 (2003), arXiv:hep-th/0305247.

\bibitem{IyerWald}
V. Iyer and R.M. Wald,
\emph{Phys. Rev. D} \textbf{50}, 846 (1994), arXiv:gr-qc/9403028;

\bibitem{IyerWald2}
R.M. Wald and A. Zoupas,
\emph{Phys. Rev. D} \textbf{61}, 084027 (2000), arXiv:gr-qc/9911095.

\bibitem{BarnichB}
G. Barnich and F. Brandt,
\emph{Nucl. Phys. B} \textbf{633}, 3 (2002), arXiv:hep-th/0111246.

\bibitem{Barnich}
 G. Barnich,
\emph{Class. Quantum Grav.} \textbf{20}, 3685 (2003), arXiv:hep-th/0301039.

\bibitem{BarnichC}
G. Barnich and G. Compere,
\emph{Phys. Rev. D} \textbf{71}, 044016 (2005). \emph{Erratum ibid} \textbf{73}, 029904 (2006), 
arXiv:gr-qc/0412029.

\bibitem{BCintegC}
G. Barnich and G. Compere,
\emph{J. Math. Phys.} \textbf{49}, 042901 (2008),
arXiv:0708.2378 [gr-qc].

\bibitem{AbbottD1}
L. F. Abbott and S. Deser,
\emph{Nucl. Phys. B} \textbf{195}, 76 (1982).

\bibitem{AbbottD2}
L. F. Abbott and S. Deser,
\emph{Phys. Lett. B} \textbf{116}, 259 (1982).

\bibitem{DeserT1}
S. Deser and B. Tekin,
\emph{Phys. Rev. Lett.} \textbf{89}, 101101 (2002),
arXiv:hep-th/0205318.

\bibitem{DeserT2}
S. Deser and B. Tekin,
\emph{Phys. Rev. D} \textbf{67}, 084009 (2003),
arXiv:hep-th/0212292.

\bibitem{Obukhov}
Y.N. Obukhov and D. Puetzfeld,
\emph{Phys. Rev. D} \textbf{90}, 024004 (2014),
arXiv:1405.4003 [gr-qc].

\bibitem{Petrov}
A.N. Petrov and R.R. Lompay,
\emph{Gen. Rel. Grav.} \textbf{45}, 545 (2013),
arXiv:1211.3268 [gr-qc].

\bibitem{ACOlea}
R. Aros, M. Contreras, R. Olea, R. Troncoso and J. Zanelli,
\emph{Phys. Rev. Lett.} \textbf{84}, 1647 (2000),
arXiv:gr-qc/9909015.

\bibitem{KimKY}
W. Kim, S. Kulkarni and S.H. Yi,
\emph{Phys. Rev. Lett.} \textbf{111}, 081101 (2013),
arXiv:1306.2138 [hep-th].

\bibitem{Papadim}
I. Papadimitriou and K. Skenderis,
\emph{JHEP} \textbf{0508}, 004 (2005),
arXiv:hep-th/0505190.

\bibitem{Holland}
S. Hollands, A. Ishibashi and D. Marolf,
\emph{Class. Quantum Grav.} \textbf{22}, 2881 (2005),
arXiv:hep-th/0503045.

\bibitem{BKraus}
V. Balasubramanian and P. Kraus,
\emph{Commun. Math. Phys.} \textbf{208}, 413 (1999),
arXiv:hep-th/9902121.


\bibitem{CiteHJPY}
S. Hyun, J. Jeong, S.A Park and S.H. Yi,
\emph{Phys. Rev. D} \textbf{90}, 104016 (2014),
arXiv:1406.7101 [hep-th].

\bibitem{JJPeng}
J.J. Peng, \emph{Eur. Phys. J. C} \textbf{74}, 3156 (2014),
arXiv:1407.4875 [gr-qc].

\bibitem{CiteChernS}
W. Kim, S. Kulkarni and S.H. Yi,
\emph{Phys. Rev. D} \textbf{88}, 124004 (2013),
arXiv:1310.1739 [hep-th].

\bibitem{MBGhass}
M. Bravo-Gaete and M. Hassa\"{i}ne,
\emph{Phys. Rev. D} \textbf{91}, 064038 (2015),
arXiv:1501.03348 [hep-th].

\bibitem{ABBGCHJ}
E. Ay\'{o}n-Beato, M. Bravo-Gaete, F. Correa, M. Hassa\"{i}ne, M.M. Ju\'{a}rez-Aubry and J. Oliva,
\emph{Phys. Rev. D} \textbf{91}, 064006 (2015),
arXiv:1501.01244 [gr-qc].

\bibitem{Wuli}
S.Q. Wu and S.L. Li, \emph{Phys. Lett. B} \textbf{746} 276 (2015),
arXiv:1505.00117 [hep-th].

\bibitem{Setare}
M.R. Setare and H. Adami,
\emph{Phys. Lett. B} \textbf{744}, 280 (2015),
arXiv:1504.01660 [gr-qc].

\bibitem{HyunJP}
S. Hyun, J. Jeong, S.A Park and S.H. Yi,
\emph{Phys. Rev. D} \textbf{91}, 064052 (2015),
arXiv:1410.1312 [hep-th].

\bibitem{CiteLifBH}
Y. Gim, W. Kim and S.H. Yi,
\emph{JHEP} \textbf{1407}, 002 (2014),
arXiv:1403.4704 [hep-th].



\bibitem{IWaldC}
V. Iyer and R.M. Wald,
\emph{Phys. Rev. D} \textbf{52}, 4430 (1995),
arXiv:gr-qc/9503052.

\bibitem{GaoWaldC}
S.J. Gao and R.M. Wald,
\emph{Phys. Rev. D} \textbf{64}, 084020 (2001),
arXiv:gr-qc/0106071.

\bibitem{TPadman}
T. Padmanabhan,
\emph{Rept. Prog. Phys.} \textbf{73}, 046901 (2010),
arXiv:0911.5004 [gr-qc].

\bibitem{Rogatko}
M. Rogatko,
\emph{Phys. Rev. D} \textbf{72}, 074008 (2005), arXiv:hep-th/0509150;
\emph{Erratum ibid} \textbf{72}, 089901 (2005).

\bibitem{BarnichSca}
G. Barnich, arXiv:gr-qc/0211031.

\bibitem{ComMuNI}
G. Compere, K. Murata and T. Nishioka,
\emph{JHEP} \textbf{0905}, 077 (2009),
arXiv:0902.1001 [hep-th].

\bibitem{ComperH}
G. Compere,
\emph{Phys. Rev. D} \textbf{75}, 124020 (2007),
arXiv:hep-th/0703004.

\bibitem{JingHuang}
J.L. Jing and S.L. Wang, \emph{Phys. Rev. D} \textbf{65}, 064001 (2002),
arXiv:gr-qc/0110037.

\bibitem{BoseNai}
S. Bose and T.Z. Naing, \emph{Phys. Rev. D} \textbf{60}, 104027 (1999),
arXiv:hep-th/9911070.

\bibitem{KillYano}
T. Houri and K. Yamamoto,
\emph{Class. Quantum Grav.} \textbf{30}, 075013 (2013),
arXiv:1212.2163 [gr-qc].

\bibitem{Sheykhi}
A. Sheykhi, \emph{Phys. Rev. D} \textbf{76}, 124025 (2007),
arXiv:0709.3619 [hep-th].

\bibitem{SheyDHen}
A. Sheykhi, M.H. Dehghani and S.H. Hendi,
\emph{Phys. Rev. D} \textbf{81}, 084040 (2010),
arXiv:0912.4199 [hep-th].

\bibitem{ChaGoSo}
C. Charmousis, B. Gouteraux and J. Soda,
\emph{Phys. Rev. D} \textbf{80}, 024028 (2009),
arXiv:0905.3337 [gr-qc].

\bibitem{CaiJS}
R.G. Cai, J.Y. Ji and K.S. Soh,
\emph{Phys. Rev. D} \textbf{57}, 6547 (1998),
arXiv:gr-qc/9708063.

\bibitem{CHMann}
K.C.K. Chan, J.H. Horne and R.B. Mann,
\emph{Nucl. Phys. B} \textbf{447}, 441 (1995),
arXiv:gr-qc/9502042.

\bibitem{SQWu}
S.Q. Wu,
\emph{Phys. Rev. D} \textbf{83}, 121502 (2011),
arXiv:1108.4157 [hep-th].

\bibitem{KunzML}
J. Kunz, D. Maison, F. Navarro-Lerida and J. Viebahn,
\emph{Phys. Lett. B} \textbf{639}, 95 (2006),
arXiv:hep-th/0606005.


\bibitem{HenMTZ}
M. Henneaux, C. Martinez, R. Troncoso and J. Zanelli,
\emph{Annals Phys.} \textbf{322}, 824 (2007),
arXiv:hep-th/0603185.

\bibitem{LuPaPo}
H. Lu, Y Pang and C.N. Pope,
\emph{JHEP} \textbf{1311}, 033 (2013),
arXiv:1307.6243 [hep-th].

\bibitem{LiuLu}
H.S. Liu and H. Lu,
\emph{Phys. Lett. B} \textbf{730}, 267 (2014),
arXiv:1401.0010 [hep-th].

\bibitem{ConfTofDT}
S. Deser and B. Tekin, Class. Quant. Grav. \textbf{23}, 7479 (2006),
arXiv:gr-qc/0609111.

\bibitem{JJPengHornCC}	
J.J. Peng, Phys. Lett. B \textbf{752}, 191 (2016),
arXiv:1511.06516 [gr-qc].


\bibitem{MaPadma1}
B.R. Majhi and T. Padmanabhan,
\emph{Phys. Rev. D} \textbf{85}, 084040 (2012),
arXiv:1111.1809 [gr-qc].

\bibitem{MaPadma2}
B.R. Majhi and T. Padmanabhan,
\emph{Phys. Rev. D} \textbf{86}, 101501 (2012),
arXiv:1204.1422 [gr-qc].

\end{thebibliography}
\end{document}